\documentclass[aps,preprint,prd,showpacs,nofootinbib]{revtex4}

\usepackage{latexsym}
\usepackage{amsmath}
\usepackage{graphicx}
\usepackage{subfigure}
\usepackage{dcolumn}
\usepackage{bm}
\usepackage{amssymb}
\usepackage{latexsym}
\usepackage{ulem}
\usepackage{color}
\usepackage[colorlinks,linkcolor=magenta,anchorcolor=cyan,citecolor=blue]{hyperref}

\def\be{\begin{equation}}
\def\ee{\end{equation}}
\def\ba{\begin{eqnarray}}
\def\ea{\end{eqnarray}}

\begin{document}

\title{Single field slow-roll inflation with step uplift to $n_s=1$}

\author{Hao-Shi Yuan$^{1,4,5}$\footnote{yuanhaoshi24@mails.ucas.ac.cn}}

\author{Ze-Yu Peng$^{2,3}$\footnote{pengzeyv@mail.ustc.edu.cn}}

\author{Yun-Song Piao$^{1,2,3,4}$\footnote{yspiao@ucas.ac.cn}}

\affiliation{$^1$ School of Fundamental Physics and Mathematical Sciences,
Hangzhou Institute for Advanced Study, UCAS, Hangzhou 310024, China}

\affiliation{$^2$ School of Physical Sciences, University of Chinese
Academy of Sciences, Beijing 100049, China}

\affiliation{$^3$ International Center for Theoretical Physics Asia-Pacific, Beijing/Hangzhou, China}

\affiliation{$^4$ Institute of Theoretical Physics, Chinese Academy of Sciences, P.O. Box 2735, Beijing 100190, China}

\affiliation{$^5$ University of Chinese Academy of Sciences, Beijing 100190, China}

\begin{abstract}

The early dark energy resolution of Hubble tension seems to be
suggesting a scale-invariant Harrison-Zeldovich spectrum of
primordial scalar perturbation, i.e. $n_s=1$ ($|n_s-1|\sim {\cal
O}(0.001)$) for $H_0\sim 73$km/s/Mpc. In this work, we propose a
possibility to acquire $n_s=1$ in single field slow-roll models of
inflation. In our consideration, the potential of inflaton during
inflation still preserve the shape of well-known single field
inflation models in deep slow-roll region, but inflation ends
suddenly due to a large step of inflaton potential. In particular,
we investigate the implication of our scheme for chaotic inflation
and Starobinski inflation, and show how they can be compatible
with the observation for $n_s=1$.

\end{abstract}
\maketitle

\section{Introduction}

It is well known that inflation is the current paradigm of early
universe
\cite{Guth:1980zm,Linde:1981mu,Albrecht:1982wi,Starobinsky:1980te,Mukhanov:1981xt,Linde:1983gd},
which predicts (nearly) scale-invariant scalar perturbation, as
well as primordial gravitational waves (GW). In popular single
field slow-roll inflation models, the spectral index $n_s$ of
primordial scalar perturbation follows
\cite{Mukhanov:2013tua,Roest:2013fha,Garcia-Bellido:2014gna,Creminelli:2014nqa},
\be n_s-1 = -{O(1)\over N_*}\label{ns-N}\ee in large-$N_*$ limit,
where $N_* = \int Hdt$ is the efolds number before the end of
inflation. The perturbation modes corresponding to cosmic
microwave background (CMB) fluctuations exit the horizon at about
$N_*\sim 60$ efolds. Recently, based on standard $\Lambda$CDM
model the Planck collaboration obtains $n_s\simeq 0.965$
\cite{Planck:2018vyg}, which is consistent with (\ref{ns-N}).

However, this result for $n_s$ depends on our understanding about
the physics before recombination. In the past ten years, a
significant conflict has emerged between the Hubble constants
inferred from the Planck collaboration \cite{Planck:2018vyg} using
their cosmic microwave background (CMB) data based on $\Lambda$CDM
and the SH0ES collaboration \cite{Riess:2021jrx} using
Cepheid-calibrated supernovas. This tension, now standing at
$\gtrsim 5\sigma$, constitutes one of the major challenges to the
standard cosmological model and might hint the new physics beyond
$\Lambda$CDM, see
Refs.\cite{DiValentino:2021izs,CosmoVerseNetwork:2025alb} for
reviews and recent developments. In the Early Dark Energy (EDE)
resolution \cite{Poulin:2018cxd,Smith:2019ihp} of
the Hubble Tension,
the Hubble constant is lifted to $H_0\gtrsim 72$km/s/Mpc through
the suppression of sound horizon. It is also possible that the EDE
potential has an anti-de Sitter (AdS) well
\cite{Ye:2020btb,Jiang:2021bab,Ye:2020oix,Wang:2022jpo}, such an
AdS-EDE can further lift $H_0$ to $H_0\sim 73$km/s/Mpc. It has
been found that the bestfit values of cosmological parameters
acquired assuming $\Lambda$CDM must shift accompanied with $\Delta
H_0$, as a result if $H_0\sim 73$ (compatible with the SH0ES
result), we have $n_s=1$ ($|n_s-1|\sim {\cal O}(0.001)$)
\cite{Ye:2021nej,Jiang:2022uyg}\footnote{The possibilities of
$n_s=1$ in different resolutions of the Hubble tension have been
also investigated in
Refs.\cite{DiValentino:2018zjj,Giare:2022rvg,Calderon:2023obf,Giare:2024akf}},
see also
\cite{Smith:2022hwi,Jiang:2022qlj,Jiang:2023bsz,Peng:2023bik,Wang:2024dka,Wang:2024tjd},
and recent \cite{Peng:2025tqt} with ACT, SPT data and DESI DR2.

Therefore, it seems that potential resolution of the Hubble
tension might be pointing to a scale-invariant Harrison-Zeldovich
spectrum of primordial scalar perturbation. Recently, relevant
inflation models has been intensively investigated. $|n_s-1|\sim
{\cal O}(0.001)$ can be implemented in single field slow-roll
model but with the exit through waterfall instabilty of another
field (see hybrid inflation), e.g.recent
\cite{Kallosh:2022ggf,Ye:2022efx,Braglia:2022phb}\footnote{It has
been observed that multiple inflation generally makes the spectrum
redder \cite{Piao:2006nm,Giare:2023kiv,Felegary:2024erb}, not
helps to the shift of $n_s$ to $n_s=1$. }, or nonminimal
derivative coupling to gravity \cite{Fu:2023tfo,Fu:2025ciy}.

In this paper, we present the possibility that acquiring $n_s=1$
in single field slow-roll model but without the exit through
waterfall instabilty of another field. As inspired by recent
Ref.\cite{Ye:2022efx}, inflation lasts about 60 efolds number dose
not necessarily require $N_*=60$. In our uplift of $n_s\simeq
0,965$ to $n_s=1$, the potential of inflaton $\phi$ during
inflation still preserve the shape of well-known single field
inflation models in deep slow-roll region in which $N_*\gg \Delta
N\simeq 60$, but inflation ends suddenly due to a large step of
inflaton potential. In particular, we investigate the implication
of our scheme for chaotic inflation and Starobinski inflation.


\section{Hint of EDEs for $n_s=1$}


In pre-recombination EDE models, an additional dark energy
component is non-negligible only for a short epoch before the
recombination without altering the recombination process and the
subsequent evolution of the $\Lambda$CDM universe.

The angular scale fixed by CMB observation ($D_{A}$ is the angular
diameter distance to the last scattering surface) equals to
\begin{align}
  \theta=\frac{r_{s}}{D_{A}} \sim r_{s}H_{0}.
\end{align}
As a result, the lower sound horizon $r_{s}$ corresponds to a
higher $H_{0}$. In AdS-EDE \cite{Ye:2020btb}, since the potential
has an AdS well, which makes EDE dilutes away faster before the
recombination, a larger EDE fraction and so higher $H_0\sim
73$km/s/Mpc can be achieved without spoiling fit to
Planck+BAO+Pantheon dataset.

\begin{figure}[htbp]
\includegraphics[width=.80\textwidth]{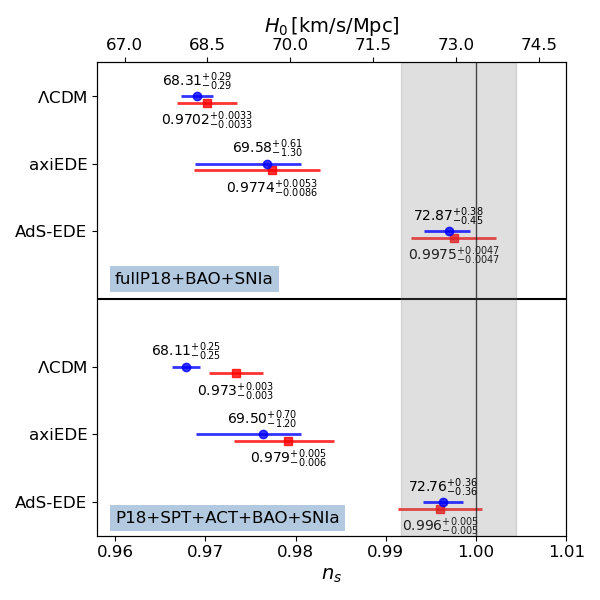}
\caption{$n_{s}-H_{0}$ relation. The data shown are sourced from
\cite{Peng:2025tqt}, where the full Planck data
\cite{Planck:2019nip}, DESI DR2 BAO data \cite{DESI:2025zgx} and
Type Ia supernovas dataset Patheon \cite{Scolnic:2021amr} are
combined for the upper panel, while the large-scale Planck 2018
data \cite{Planck:2019nip}, ACT
DR6\cite{AtacamaCosmologyTelescope:2025blo,AtacamaCosmologyTelescope:2025nti},
SPT-3G D1\cite{SPT-3G:2025bzu,Balkenhol:2024sbv}, together with
BAO and SNIa, are responsible for the lower panel. Gray band
represents the recent SH0ES result $H_{0}=73.04\pm1.04$ km/s/Mpc
\cite{Riess:2021jrx}, and black solid line marks $n_{s}=1$.}
\label{fig1}
\end{figure}

The shift of spectral index $n_{s}$ satisfies an universal
$n_s$-$H_0$ scaling relation \cite{Ye:2021nej}:
\begin{align}
  \delta n_{s} \simeq 0.8(1-\alpha) \frac{\delta H_{0}}{H_{0}},
\label{variation relation}
\end{align}
where $\alpha \simeq 0.5$ for the Planck+BAO+Pantheon dataset and
$\alpha \simeq 0.7$ for the Planck+(earlier ACT+SPT)+BAO+Pantheon
dataset \cite{Peng:2025tqt}. Therefore, the $\Lambda$CDM model
with pre-recombination EDE predicts the scale-invariant
Harrison-Zel'dovich spectrum $n_{s}\simeq 1$ for $H_0\sim
73$km/s/Mpc, resulting in profound implications for the inflation
and primordial universe, see e.g.recent
\cite{Kallosh:2022ggf,Ye:2022efx,Jiang:2023bsz,Giare:2024akf,Takahashi:2021bti,Braglia:2022phb}.
Combined latest observational datasets, like
Planck+SPT-3G+ACT-DR6+BAO+Pantheon, support this conclusion, with
relevant results illustrated in Fig.\ref{fig1}.

\section{Inflation with step uplift to $n_s=1$}
\label{secIII}


The end of inflation is marked with the violation of the slow-roll
condition, $\varepsilon_{V}(\phi=\phi_{e}) \simeq 1$, which is
defined by
\begin{align}
  \varepsilon_{V}(\phi) = \frac{1}{2}\left(\frac{V_{\phi}}{V}\right)^2
\end{align}
where $V_{\phi}\equiv dV/d\phi$, in units where
$M_{\mathrm{Pl}}=1/\sqrt{8\pi G}$ is set to unity, and the other
potential slow-roll parameter is
$\eta_{V}=\frac{V_{\phi\phi}}{V}$. Physical quantities
corresponding to the original and modified potentials are
distinguished by the subscripts $o$ and $s$ respectively.

\begin{figure}[htbp]
\includegraphics[width=.8\textwidth]{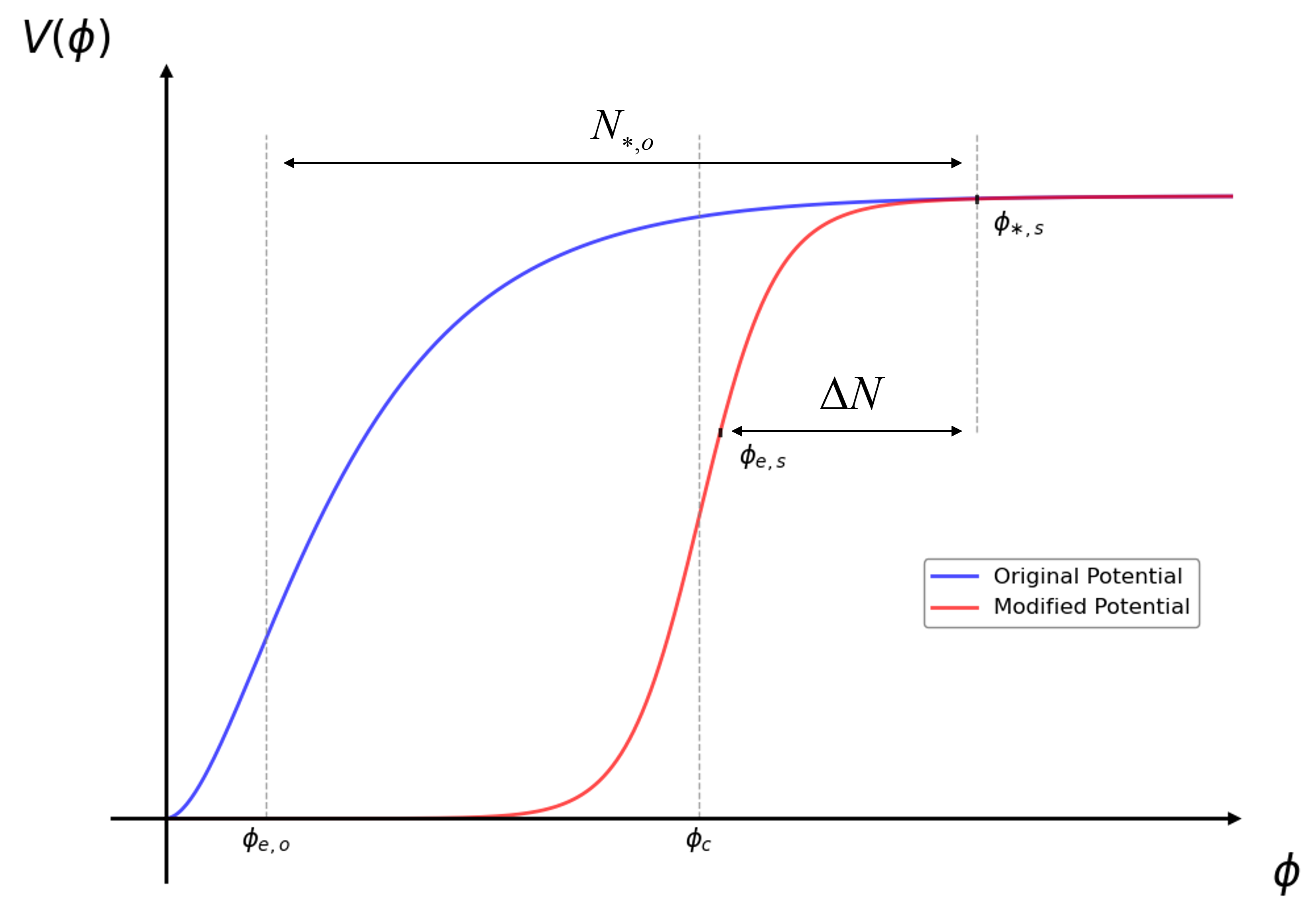}
\caption{The slow-roll potential and its step-modified version. In
both cases, inflation ended at the position where $\varepsilon_{V}
= 1$, corresponding to $\phi_{e,o}$ and $\phi_{e,s}$ respectively.
The perturbation mode that exited the horizon at $N_{\ast,o}\gg
60$ remains far outside the current observable Hubble radius.
However, with the abrupt step-caused exit of inflation, this mode
can be regarded as the CMB window, and accordingly gives $n_{s}\to
1$.} \label{inflationstep}
\end{figure}

In most single field slow-roll model of inflation, $n_s$ follows
the relation (\ref{ns-N}). According to (\ref{ns-N}),
$N_{\ast}\approx 60$ is incompatible with results of $n_s$ in
$H_0\gtrsim 72$km/s/Mpc cosmologies, which requires larger
$N_{\ast}$ to satisfy $n_{s} \approx 0.990$ at least. However,
inflation lasts $\Delta N \approx 60$ does not necessarily require
$N_{\ast,o} \approx 60$ in (\ref{ns-N}), in particular if the deep
slow-roll phase meets a step-like break point of potential so that
inflation abruptly ended, see Fig.\ref{inflationstep}.


In corresponding scenario, the efolds number $N_{\ast,o}$ in
original popular slow-roll models, such as chaotic inflation
\cite{Linde:1983gd}, Starobinski inflation
\cite{Starobinsky:1980te}, brane inflation
\cite{Dvali:1998pa,Burgess:2001fx,Kachru:2003sx}, can be
decomposed by, using slow-roll approximation,
\begin{align}
N_{\ast,o} \approx
\int_{\phi_{e,o}}^{\phi_{\ast,s}}\frac{1}{\sqrt{2\varepsilon_{V,o}}}
d\phi \equiv \Delta N + N_{o}(\phi_{e,o} \to \phi_{e,s}) + \Delta
N_{step}
\end{align}
where $\Delta N=
\int_{\phi_{e,s}}^{\phi_{\ast,s}}d\phi/\sqrt{2\varepsilon_{V,s}}
\approx 60$ accounts for the efolds number of slow-roll inflation
with the step-modified potential, and
\begin{align}
N_{o}(\phi_{e,o} \to \phi_{e,s}) =
\int_{\phi_{e,o}}^{\phi_{e,s}}\frac{1}{\sqrt{2\varepsilon_{V,o}}}
d\phi,
\end{align}
corresponds to the efolds number accumulated under the original
potential as the inflaton rolls from the step-caused end point to
the original end point. It can be achieved $\mathcal{O}(10^{2})$,
playing a crucial role in raising the spectral index to
$n_s\gtrsim 0.99$. $\Delta N_{step}$ represents the possible
impact of step on the efolds number. The validity of the relation
between tilt and efolds (\ref{ns-N}) requires that it must be
negligibly small compared to $N_{o}(\phi_{e,o} \to \phi_{e,s})$.
Its contribution is sensitive to the width parameter of the step
$d$ we choose. In particular when the step is sharp enough, we
have $\Delta N_{step}\ll {\cal O}(1)$. As a result we can neglect
it. Thus we can acturally have $n_{s}$ close to 1 by pushing the
position of the step to a deeper slow-roll region of the original
potential so that $N_{\ast,o}\gg \Delta N \approx 60$ while ending
inflation by the sharp step at $N_{\ast,o}-60$.


\section{Applications to some popular models}
\label{secIV}

In our uplift model of $n_{s}\approx 0.965$ to $n_{s}\approx 1$,
we consider the inflation potential as
\begin{align}
V_s(\phi)=f_{step}V_{o}(\phi)
\label{step potential}
\end{align}
with \be
f_{step}=\frac{1}{2}\left[1+\tanh\left(\frac{\phi-\phi_{c}}{d}\right)\right].\ee
where $V_{o}(\phi)$ is the single field slow-roll potential. In
(\ref{step potential}), when $\phi\gg \phi_c$,
$V_s(\phi)=V_{o}(\phi)$, the evolution of inflaton follows that
with the corresponding single field slow-roll potential, while
when $\phi\simeq\phi_c$, the potential become abrupt\footnote{In
our consideration, when $\phi\ll\phi_c$, we have $V_s(\phi)=0$.
However, it is also possible that $V_s(\phi)<0$, i.e. the universe
went through an (or multiple) AdS phase and then climb up into
$\Lambda$CDM evolution, see
e.g.\cite{Li:2019ipk,Wang:2025dtk,Piao:2005ag}.} so that the
inflation can rapidly desist.

It is well-known that the corresponding step, possibly inspired by
the supergravity theory \cite{Adams:2001vc}, can result in
oscillations in the power spectrum of primordial perturbations,
which might improve the fit to the CMB data, or explain certain
CMB anomalies
e.g.\cite{Adams:2001vc,Covi:2006ci,Chen:2006xjb,Hamann:2007pa,Chen:2008wn,Mortonson:2009qv,Dvorkin:2009ne,Hazra:2010ve,Liu:2010dh,Adshead:2011jq,Adshead:2012xz,Adshead:2013zfa,Benetti:2012wu,Benetti:2013cja,Miranda:2013wxa,Cai:2015xla,Miranda:2015cea,GallegoCadavid:2016wcz,Fard:2017oex}.
Recently, the potentials with a sharp step on small scales have
been also researched to significantly amplify the power spectrum
within a certain range of scales, a compelling mechanism for
generating primordial black holes
\cite{Kefala:2020xsx,Dalianis:2021iig,Inomata:2021tpx,Inomata:2021uqj,Rojas:2022iho,Mastache:2023cha,Thomas:2023poh,Thomas:2024ezg,Mohammadi:2025avz}.
However, what we consider is a large step enough to suddenly end
slow-roll inflation. Here, the primordial spectrum of scalar
perturbation inherit that of original potential $V_{o}(\phi)$,
i.e.(\ref{ns-N}), and the step only affect $N_*$ so $n_s-1$.


\subsection{Monomial inflation}

\begin{figure}[htbp]
    \centering
    \includegraphics[width=0.8\textwidth]{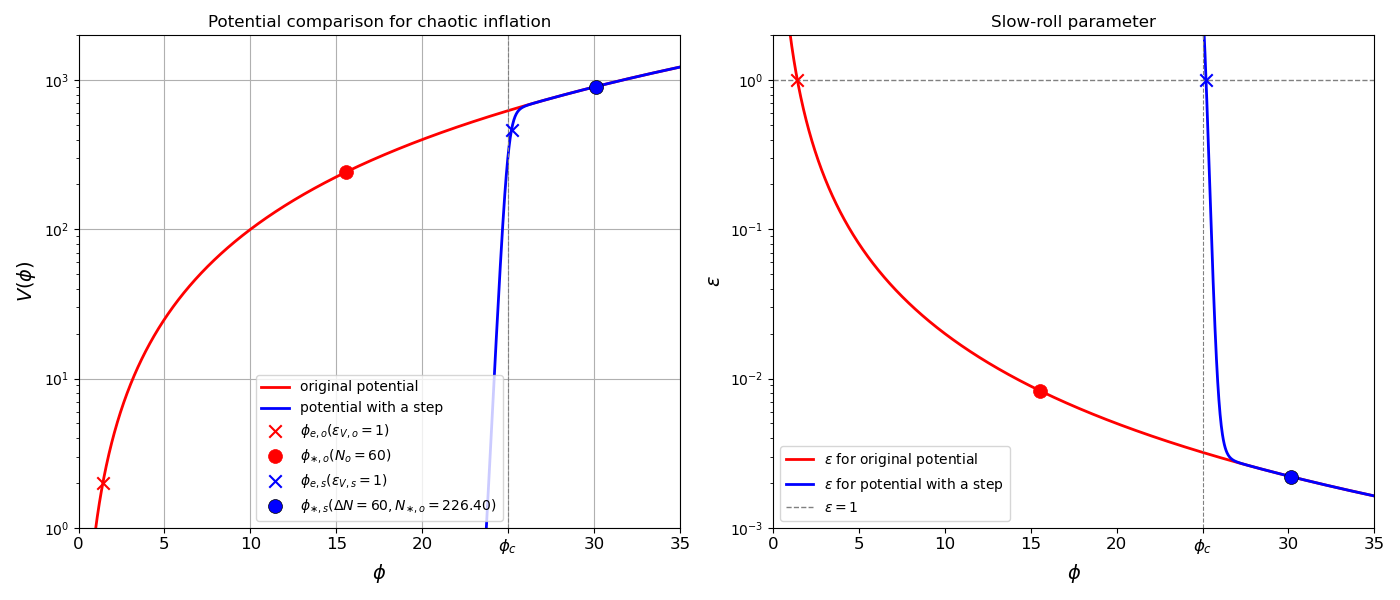}
    \vspace{0.5cm}
    \includegraphics[width=0.8\textwidth]{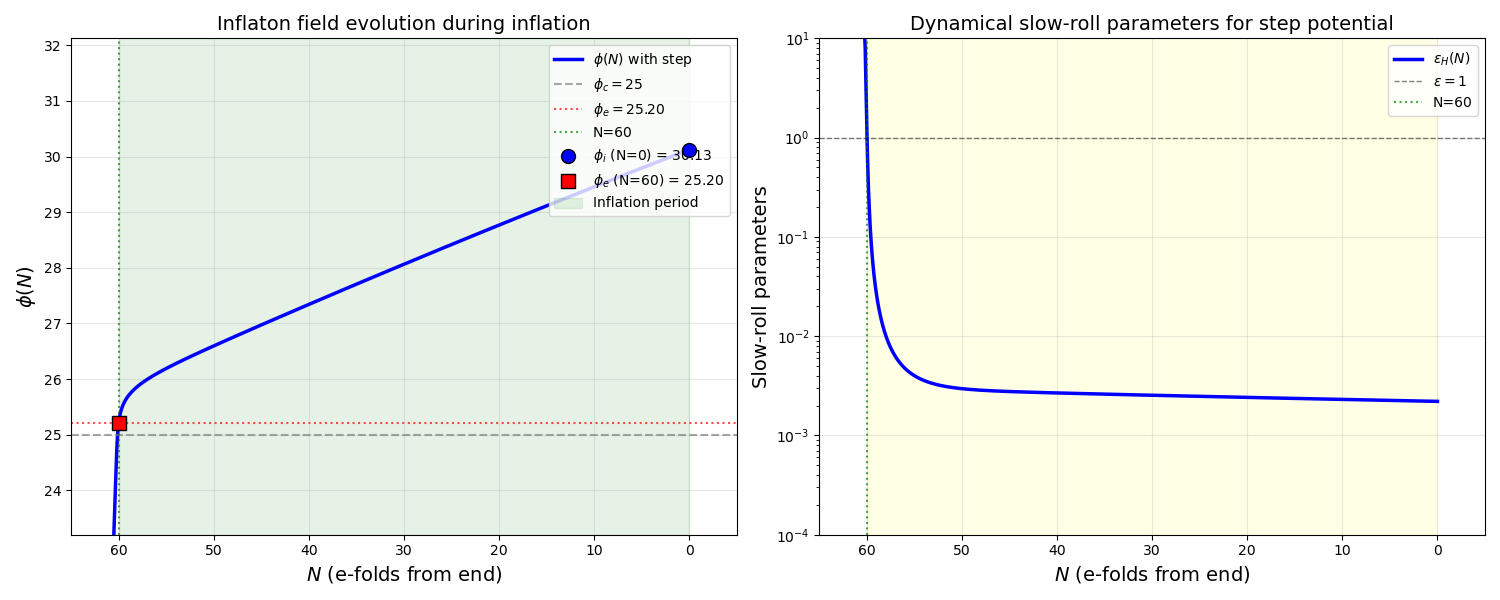}
\caption{Comparison of chaotic inflation with and w/o the step.
The parameters of the step are $\phi_{c}=25,\ d=0.4$. Top panels:
the shapes, potential slow-roll parameters of both models. The
inflation happened from $\phi_{\ast}=30.12$ to $\phi_{s}=25.20$,
predicting a larger spectral index $n_{s}\simeq 0.991$. Bottom
panels: the evolution of the inflaton $\phi$ and the slow-roll
parameter $\varepsilon_{V},\ \eta_{V}$ with respect to efolds
number $N$.}
    \label{chaotic n=2}
\end{figure}

\begin{figure}[htbp]
\includegraphics[width=0.8\textwidth]{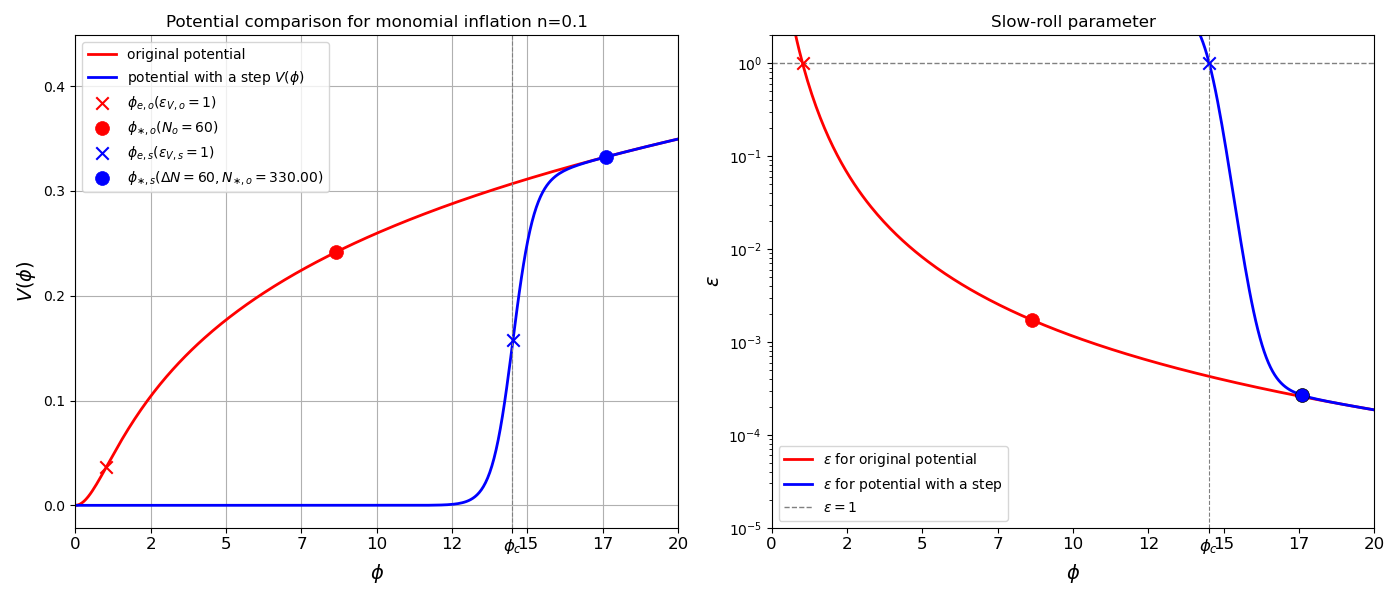}
\caption{Comparison of monomial inflation with (blue) and w/o the
step (red). Here we choose $n=0.1,\ \phi_{c}=14.5,\ d=0.7$. Left
panel: Intuitive illustration of the effect of the step in the
shape of the potential. Right panel: The evolution of slow-roll
parameter $\varepsilon$ as the function of inflaton $\phi$.}
\label{monomial n=0.1}
\end{figure}

In usual monomial inflation, $V_{0}(\phi)\sim \phi^{n}$, we have
\begin{align}
  n_{s}-1\simeq -\frac{n/2+1}{N_{\ast}},\quad r\simeq \frac{4n}{N_{\ast}}
\end{align}
which includes the well-known chaotic inflation
\cite{Linde:1983gd} for $n=2$, and the monodromy inflation for
$n=2/3$, $1$ or smaller $n$
\cite{McAllister:2008hb}.

The chaotic inflation predicts tensor-to-scalar ratio
$r\simeq0.13$, far exceeds the BICEP/Keck upper limit $r<0.036$
\cite{BICEP:2021xfz} and is strongly disfavored. However, with our
step modification (\ref{step potential}), it can be seen that the
inflation occurs under the slow-roll condition, $\epsilon_{V}\ll
1$, see Fig.\ref{chaotic n=2}, and then the inflaton rapidly exits
the slow-roll phase due to the steepness of the step, causing
inflation to end abruptly, however, the slow-roll phase can be
pushed towards the deeper slow-roll region of the potential
unaffected by the step, to achieve $\Delta N=60$ efolds. It is
observed that when $N_{\ast,o}>200$ in (\ref{ns-N}) we can have a
extremely nearly scale-invariant Harrison-Zeldovich spectrum
$0.99<n_{s}<1 $, the tensor-to-scalar ratio $r$ is now $r\simeq
0.035$, which is consistent with the results in EDE cosmologies at
$\sim 2\sigma$ level,
e.g.\cite{Ye:2022afu,Wang:2024tjd}\footnote{It should be mentioned
that the upper on $r$ in EDE cosmologies is slightly lower.}.

It is also interesting to see the $n=0.1$ case. The parameter set
of the step we adopt is $\phi_{c}=14.5,\ d=0.7$. The comparative
evolution of the potentials and slow-roll parameters $\varepsilon$
with and without step modification are showed in Fig.\ref{monomial
n=0.1}. The corresponding spectral index is still $n_{s}>0.99$ but
the tensor-to-scalar ratio of $r=0.0043$ has been significantly
suppressed.


\subsection{Starobinsky inflation}

\begin{figure}[htbp]
\includegraphics[width=0.8\textwidth]{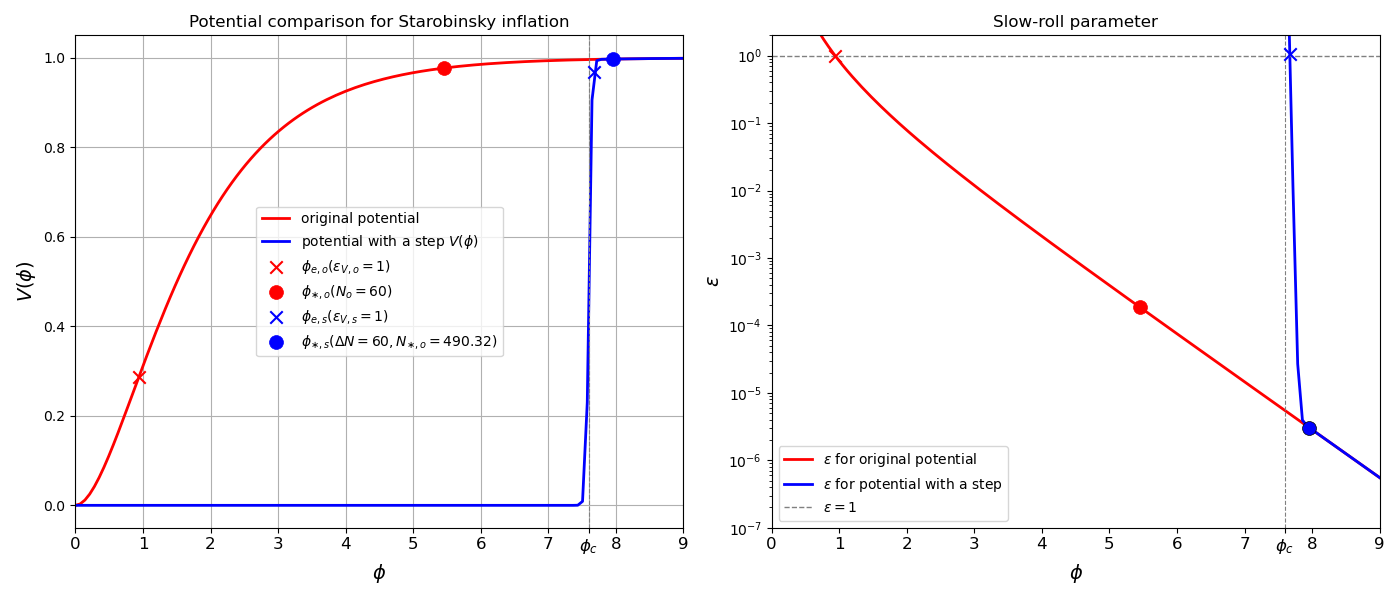}
\caption{Starobinsky inflation with the step.} \label{fig:starobinsky}
\end{figure}

In Starobinsky model \cite{Starobinsky:1980te}, the effective
potential is given by
\begin{align}
  V_{o}(\phi)\sim \left(1-e^{-\sqrt{2/3}\phi}\right)^{2},
\label{starobinsky}
\end{align}
with observational predictions:
\begin{align}
  n_{s}-1=-\frac{2}{N_{\ast}},\quad r=\frac{12}{N_{\ast}^{2}},
\label{ns-N alpha-attractor}
\end{align}
which corresponds to $\alpha=1$ in $\alpha$-attractor inflation
\cite{Kallosh:2013yoa,Kallosh:2013hoa,Kallosh:2013maa}. The
Starobinsky inflation predicts $n_{s}=0.967$ and $r=0.0033$, in
agreement with the observational results based on the $\Lambda$CDM
model. In our step uplift, $N_{\ast,o} \gtrsim 500 \gg \Delta N$
and inflation ended at $N_{\ast,o} -60$. The parameters are chosen
as $\phi_{c}=7.6$ and $d=0.04$, making the step sufficiently sharp
to ignore the $\Delta N_{step}$  and thus
$n_{s}\simeq 1-2/N_{\ast,o}\gtrsim 0.996$, highly consistent with
a scale-invariant spectrum required by Planck+SPT+ACT, see Fig.\ref{fig:starobinsky},
and $r\simeq 12/N_{\ast,o}^{2}\lesssim 4.8 \times 10^{-5}$, which
currently is also consistent with the upper bound on $r$.

\section{Discussion}

There is a hint that the potential resolution of Hubble tension
might be pointing to a scale-invariant Harrison-Zeldovich spectrum
of primordial scalar perturbation, i.e. $n_s=1$ for $H_0\sim
73$km/s/Mpc. This result dose not favor current well-known models
of inflation satisfying (\ref{ns-N}), though they are actually
well motivated. In this work, we propose a scheme to lift $n_s$
predicted by these well-known slow-roll inflation models to
$n_s=1$. In our consideration, inflation can abruptly end when
inflaton meets a step-like break point of potential but before
that time it is still at a deep slow-roll region. As examples, we
show how chaotic inflation and Starobinski inflation can be
compatible with the observation for $n_s=1$. This in certain sense
implies that the observational shift of $n_s$ might not
essentially affect our current favour on models of inflation.

The step-like potential have been used widely in inflation models.
There might be some steps with different vertical drops at
different scale, one at large scale responsible for the possible
oscillations in primordial scalar perturbation, one for other
phenomenologies at small scale, e.g.\cite{Li:2020cjj} for explain
recent PTA GW, and one for the end of inflation. Thus our work not
only make well-known single field slow-roll inflation models more
adaptable when they are confronted with changeable $n_s$, but also
highlight the role of step-like potential which might result in
lots of observable signals to be explored.

\textbf{Acknowledgments} This work is supported by NSFC,
No.12475064, National Key Research and Development Program of
China, No. 2021YFC2203004, and the Fundamental Research Funds for
the Central Universities.


\end{document}